\documentclass[11pt,a4paper]{article}

\usepackage{jheppub}
\pdfoutput=1
\usepackage{amssymb}
\usepackage{amsbsy}
\usepackage{amsfonts}
\usepackage{amsmath}
\usepackage{graphicx}
\usepackage{amssymb}
\usepackage{xcolor}
\usepackage{slashed} 
\usepackage{setspace}
\usepackage{amstext}    
\usepackage{wasysym}
\usepackage{hyperref}
\usepackage{accents} 
\usepackage{mathrsfs}
\usepackage{mdframed} 
\usepackage{booktabs} 
\usepackage{multirow}
\usepackage{subfig}
\usepackage{array,mathtools}
\newcolumntype{L}[1]{>{\raggedright\arraybackslash}p{#1}}

\newcommand{\abs}[1]{\left\vert#1\right\vert}

\newcommand{\esp}[1]{\left[#1\right]}
\newcommand{\corc}[1]{\left(#1\right)}
\newcommand{\llav}[1]{\left\{#1\right\}}

\DeclareMathAlphabet{\mathpzc}{OT1}{pzc}{m}{it}

\newcommand{\ds}[1]{{\displaystyle #1 }}

\newcommand{\CNB}{\rm C\nu B}

\newcommand{\blue}[1]{\color{blue} #1 \color{black}}

\graphicspath{{figures/}}

\newcolumntype{C}{>{$}c<{$}}
\AtBeginDocument{
\heavyrulewidth=.08em
\lightrulewidth=.05em
\cmidrulewidth=.03em
\belowrulesep=.65ex
\belowbottomsep=0pt
\aboverulesep=.4ex
\abovetopsep=0pt
\cmidrulesep=\doublerulesep
\cmidrulekern=.5em
\defaultaddspace=.5em
}

\title{\blue{Impact of Beyond the Standard Model physics in the detection of the Cosmic Neutrino Background}}

\author{Mart\'in Arteaga,}
\emailAdd{martin77@if.usp.br}

\author{Enrico Bertuzzo,}
\emailAdd{bertuzzo@if.usp.br}

\author{Yuber F. Perez-Gonzalez and}
\emailAdd{yfperezg@if.usp.br}

\author{Renata Zukanovich Funchal}
\emailAdd{zukanov@if.usp.br}

\affiliation{Departamento de F\'isica Matem\'atica, Instituto de F\'isica, Universidade de S\~ao Paulo,\\
R. do Mat\~ao 1371, CEP.\ 05508-090, S\~ao Paulo, Brazil}

\abstract{We discuss the effect of Beyond the Standard Model charged
  current interactions on the detection of the Cosmic Neutrino
  Background by neutrino capture on tritium in a PTOLEMY-like
  detector. We show that the total capture rate can be substantially
  modified for Dirac neutrinos if scalar or tensor right-chiral
  currents, with strength consistent with current experimental bounds,
  are at play. We find that the total capture rate for Dirac
  neutrinos, $\Gamma_{\rm D}^{\rm BSM}$, can be between 0.3 to 2.2 of 
  what is expected for Dirac neutrinos in the Standard Model, 
  $\Gamma_{\rm D}^{\rm SM}$, so that it can be made as large as 
  the rate expected for Majorana neutrinos with only Standard Model 
  interactions. A non-negligible primordial abundance of right-handed 
  neutrinos can only worsen the situation, increasing 
  $\Gamma_{\rm D}^{\rm BSM}$ by 30 to 90\%. On the other hand, if 
  a much lower total rate is measured than what is expected for 
  $\Gamma_{\rm D}^{\rm SM}$, it may be a sign of new physics.} 
   \arxivnumber{}

\begin{document}
\maketitle

\section{Introduction}

The accidental discovery of the Cosmic Microwave Background (CMB)
radiation by Penzias and Wilson in 1965 laid the foundations for the
enormous progress in our understanding of the evolution of the
Universe. This is the oldest directly observed radiation
in the Universe, dating from the epoch of recombination, and its
precise study, carried out in the last decades by various cosmological
probes, lead to the establishment of the standard model of cosmology.
This model also predicts the existence of a Cosmic Neutrino Background
($\CNB$), a relic radiation that decoupled from matter when the
Universe was merely a second old, which is expected to have played
a crucial role in primordial nucleosynthesis and in large scale
structures formation.
 
The CMB anisotropies, an indirect imprint of the $\CNB$, have already 
offered two important constraints in connection to particle physics: 
a limit on the sum of neutrino masses and the effective number of 
neutrino species.  A confirmation of the $\CNB$ by direct detection 
using experiments on Earth would not only represent a further triumph 
of modern cosmology, but it would also constitute an unique opportunity 
to probe neutrino properties. For a long time this was believed to be 
an impossible task since relic neutrinos are expected to be non-relativistic 
today with an average momentum of about $10^{-4}$ eV. Recent developments 
have allowed to revive the old suggestion by Weinberg~\cite{Weinberg:1962zza} 
of capturing them on $\beta$-decaying nuclei, a process with no energy
threshold. In fact, a real experimental proposal, the Princeton
Tritium Observatory for Light, Early-Universe, Massive-Neutrino Yield
(PTOLEMY) experiment~\cite{Betts:2013uya} is currently assessing the
prospects for using the process $\nu +\ ^{3}{\rm H} \to\ ^{3}{\rm He}+
e^-$. The signature of $\CNB$ capture would be a peak in the final
electron spectrum at an energy $2 m_\nu$ above the $\beta$-decay
endpoint. This requires a very challenging energy resolution $\lesssim
0.1 $ eV for the final electrons to be distinguished from the
$\beta$-decay background. This has triggered interest in the community
to investigate what could potentially be learned in such
experiment~\cite{Long:2014zva,Zhang:2015wua,Chen:2015dka,deSalas:2017wtt}.

In particular, the authors of ref.~\cite{Long:2014zva} have shown how
the direct measurement of the $\CNB$ would allow to
discriminate Majorana from Dirac neutrinos, as the former would
produce a capture rate twice as large as the latter.  This is
because for non-relativistic states chirality and helicity do not
coincide, and it is helicity, not chirality, which is conserved by the
$\CNB$. Their conclusions rely on the fact that only the neutrinos
that interact weakly according to the Standard Model (SM) could be
produced and kept in thermal equilibrium before decoupling, a feature
that could be modified by new interactions or a different thermal
history~\cite{Zhang:2015wua,Chen:2015dka}.

In this paper we try to answer the following question: \emph{if
  neutrinos have new Beyond the Standard Model (BSM)
  interactions, how would this affect the relic neutrino detection
  rate in PTOLEMY-like detectors?} We implement these possible
deviations using an effective lagrangian approach.

We start in section~\ref{sec:EFT} by describing the gauge invariant
operators that we will consider and computing the rate of neutrino capture
on tritium. In section~\ref{sec:ptolemy} we introduce the experimental
resolution and describe in detail when the signal from the electron
produced in the capture can be distinguished from the electron
produced by the $\beta$-decay background. In section~\ref{sec:bounds}
we discuss the experimental bounds from $\beta$-decay on the BSM 
physics coefficients, and we show how the capture rate is modified
with respect to the standard case for various regions of the parameter
space. In section~\ref{sec:cosmo} we discuss how gravitational clustering 
or a primordial abundance of right-handed neutrinos present in the $\CNB$
today would affect our results. Finally our conclusions are drawn in 
\ref{sec:conc}. In appendix~\ref{sec:ordering} we discuss how the interplay 
between the experimental resolution and the neutrino mass ordering affect 
the possibility of distinguishing  the electron peaks due to each neutrino 
mass eigenstate.

\section{Effective lagrangian approach for the BSM neutrino interactions}\label{sec:EFT}

In the SM, the weak interactions have a purely $V-A$ Lorentz structure. Since 
the simple fact that neutrinos have a non-zero mass constitutes already an evidence for BSM physics, 
we will allow here for other possibilities. This can be done in a model independent 
fashion using an effective field theory approach. We will consider dimension-six 
operators which are SU(2)$_L\times$U(1)$_Y$ invariant, but which also include right-handed 
neutrinos~\cite{Grzadkowski:2010es,Cirigliano:2012ab,Cirigliano:2013xha}. More precisely, 
we write
\begin{align}\label{eq:d6OPer}
	\mathscr{L}_{\rm BSM}=\mathscr{L}_{\rm SM}^{(4)} + \mathscr{L}_{m_\nu}+\frac{1}{\Lambda^2}\sum_{k=1}^{12} c_k^{(6)} Q_k^{(6)},
\end{align}
where $\mathscr{L}_{\rm SM}^{(4)}$ is the dimension-four SM lagrangian, $\mathscr{L}_{m_\nu}$ is 
the neutrino mass lagrangian, which can either come from a dimension 4 operator involving right handed 
neutrinos or from the dimension 5 Weinberg operator; $\Lambda$ is the maximum energy scale at which the 
theory is still valid; and the $c_k^{(6)}$ are dimensionless coupling constants. The set of operators 
with left- and right-handed neutrinos, $Q_k^{(6)}=\{Q_k^{(6)}(\nu_L),Q_k^{(6)}(\nu_R)\}$, is given 
in table~\ref{tab:d6OPer}.
\begin{table}[t]
	\begin{tabular}{lll}
	\toprule\toprule
	\multicolumn{2}{c}{Four-fermion Operators} &  \multicolumn{1}{c}{Vertex Corrections}\\ 
	\cmidrule(lr){1-2}\cmidrule(lr){3-3}
	\multicolumn{1}{c}{$Q_{\nu_L}^{(6)}$} & \multicolumn{1}{c}{$Q_{\nu_R}^{(6)}$} &  \multicolumn{1}{c}{$Q_{\Phi}^{(6)}$} \\
	\midrule
		$Q_1=(\overline{l_L}e_R)(\overline{d_R}q_L)$ & $Q_5=(\overline{l_L}\nu_R)\varepsilon(\overline{q_L}d_R)$ & $Q_9=i(\Phi^T \varepsilon D_\mu\Phi)(\overline{u_R}\gamma_\mu d_R)$\\ 
		$Q_2=(\overline{l_L}e_R)\varepsilon(\overline{q_L}u_R)$ & $Q_6=(\overline{\nu_R}l_L)(\overline{q_L}u_R)$ & $Q_{10}=i(\Phi^T \varepsilon D_\mu\Phi)(\overline{\nu_R}\gamma^\mu e_R)$ \\
		$Q_3=(\overline{l_L}\gamma^\mu \tau^A l_L)(\overline{q_L}\gamma_\mu \tau^A q_L)$ & $Q_7=(\overline{e_R}\gamma^\mu \nu_R)(\overline{u_R}\gamma_\mu d_R)$ & $Q_{11}=(\Phi^\dagger i \overset{\text{\tiny$\longleftrightarrow$}}{D_\mu^a} \Phi)(\overline{q_L}\gamma_\mu \tau^A q_L)$\\ 
		$Q_4=(\overline{l_L}\sigma^{\mu\rho}e_R)\varepsilon (\overline{q_L}\sigma_{\mu\rho}u_R)$ & $Q_8=(\overline{l_L}\sigma^{\mu\rho}\nu_R)\varepsilon(\overline{q_L}\sigma_{\mu\rho}d_R)$  & $Q_{12}=(\Phi^\dagger i \overset{\text{\tiny$\longleftrightarrow$}}{D_\mu^a} \Phi)(\overline{l_L}\gamma^\mu \tau^A l_L)$\\ 
	\bottomrule
	\end{tabular}
	\caption{Dimension-six operators relevant for neutrino capture. Here $l_L,q_L$ are the SM lepton and quark SU(2)$_L$ doublets
	 while $u_R,d_R,e_R,\nu_R$ are the corresponding SM singlets. The SU(2)$_L$ generators are denoted with $\tau^A$ while
	 $\varepsilon_{ij}$ is the totally antisymmetric tensor with $\varepsilon_{12}=+1$. We do not include the invariant 
	 operator $(\overline{\nu_R}\sigma^{\mu\rho}l_L)(\overline{q_L}\sigma_{\mu\rho}u_R)$ in the list because it does 
	 not contribute to the relic capture.}
	\label{tab:d6OPer}
\end{table}
The terms relevant for our calculation of the BSM relic neutrino capture rate on $\beta$-decaying tritium can be obtained writing eq.~(\ref{eq:d6OPer}) in terms of mass eigenstates
\begin{align}\label{eq:lagr}
	\mathscr{L}_{\rm eff}=-\frac{G_F}{\sqrt{2}} V_{ud}\, U_{ej}\, \left\{[\bar{e}\gamma^\mu(1-\gamma^5)\nu_j][\bar{u}\gamma_\mu(1-\gamma^5)d] 
	+\sum_{l,q} \epsilon_{lq} [\bar{e}\mathscr{O}_l\nu_j][\bar{u}\mathscr{O}_q d]\right\}+{\rm h.c.},
\end{align}
where a sum over the three neutrino mass eigenstates $j=1,2,3$ is implied. The couplings $\epsilon_{lq}$, 
related to the dimensionless couplings $c_k^{(6)}$ (see ref.~\citep{Cirigliano:2012ab}), 
parametrize the BSM physics effects, with $l$ ($q$) labelling the Lorentz structure of 
the lepton (quark) current, as given by $\mathscr{O}_l$ ($\mathscr{O}_q$) in table~\ref{tab:ParOPer}. 
$V_{ud}$ and $U_{ej}$ correspond to the Cabibbo-Kobayashi-Maskawa (CKM) and 
Pontecorvo-Maki-Nakagawa-Sakata (PMNS) mixing matrices elements relevant to the process, respectively.

Equation~(\ref{eq:lagr}) can be used to calculate the neutrino absorption on tritium
$$ \nu_j +\ ^3{\rm H} \to\ ^{3}{\rm He} + e^{-}\, ,$$
in the presence of BSM interactions. To this end, we need to properly define the hadronic matrix elements involving the quark current in eq.~(\ref{eq:lagr}). Following ref.~\cite{Ludl:2016ane}, we have
\begin{align}
	\begin{aligned}
		\langle p(p_p)|\bar{u}\gamma^\mu(1\pm\gamma^5)d |n(p_n)\rangle &=  \overline{u_p}(p_p)\gamma^\mu[g_V(q^2)\pm g_A(q^2)\gamma^5]u_n(p_n),\\
		\langle p(p_p)|\bar{u}d |n(p_n)\rangle &=  g_S(q^2)\,\overline{u_p}(p_p)\,u_n(p_n),\\
		\langle p(p_p)|\bar{u}\gamma^5 d |n(p_n)\rangle &= g_P(q^2)\, \overline{u_p}(p_p)\gamma^5 u_n(p_n),\\
		\langle p(p_p)|\bar{u}\sigma^{\mu\nu}(1\pm\gamma^5)d |n(p_n)\rangle &= g_T(q^2)\, \overline{u_p}(p_p)\sigma^{\mu\nu}(1\pm\gamma^5)u_n(p_n).
	\end{aligned}
\end{align}
We have introduced the hadronic form factors $g_h(q^2)$, with $h=V,A,S,P,T$
cor\-res\-pon\-ding to the vector, axial, scalar, pseudoscalar and tensor
Lorentz structures, respectively.\footnote{Since it does not contribute to the $\CNB$ capture, we do not include 
  the weak magnetic term
  \center{$\langle p(p_p)|\bar{u} \gamma_\mu d |n(p_n)\rangle_{\rm WM} = -i\ds{\frac{g_{\rm WM}}{2 M_N}} 	\overline{u_p}
  (p_p)\sigma_{\mu\nu}(p_n-p_p)^\nu u_n(p_n).$}}
Although these form factors depend on the transferred momentum $q^2=(p_n -p_p)^2$, for the capture rate we are only interested in the $q^2 \simeq 0$ limit. 
In our numerical analysis we will use the values shown in table~\ref{tab:HadPar} \cite{Akulov:2002gh,Gonzalez-Alonso:2013ura,Bhattacharya:2015esa}.
\begin{table}[t]
\begin{center}
	\begin{tabular}{ccc}
		\toprule\toprule
          $\epsilon_{lq}$ & $\mathscr{O}_l$ & $\mathscr{O}_q$ \\ \midrule
		$\epsilon_{LL}$ & $\gamma^\mu(1-\gamma^5)$ & $\gamma_\mu(1-\gamma^5)$ \\ 
		$\epsilon_{LR}$ & $\gamma^\mu(1-\gamma^5)$ & $\gamma_\mu(1+\gamma^5)$\\
		$\epsilon_{RL}$ & $\gamma^\mu(1+\gamma^5)$ & $\gamma_\mu(1-\gamma^5)$ \\ 
		$\epsilon_{RR}$ & $\gamma^\mu(1+\gamma^5)$ & $\gamma_\mu(1+\gamma^5)$\\
		$\epsilon_{LS}$ & $1-\gamma^5$ & $1$ \\ 
		$\epsilon_{RS}$ & $1+\gamma^5$ & $1$ \\ 
		$\epsilon_{LP}$ & $1-\gamma^5$ & $-\gamma^5$ \\ 
		$\epsilon_{RP}$ & $1+\gamma^5$ & $-\gamma^5$ \\ 
		$\epsilon_{LT}$ & $\sigma^{\mu\nu}(1-\gamma^5)$ & $\sigma_{\mu\nu}(1-\gamma^5)$ \\ 
		$\epsilon_{RT}$ & $\sigma^{\mu\nu}(1+\gamma^5)$ & $\sigma_{\mu\nu}(1+\gamma^5)$\\ \bottomrule
	\end{tabular}
	\caption{Parameters and their corresponding Lorentz structures for the BSM currents considered in this work.}
	\label{tab:ParOPer}
\end{center}
\end{table}
Following the calculation of ref.~\cite{Long:2014zva}, the capture cross section for a 
neutrino mass eigenstate $j$, with helicity $h_j=\pm 1$ and velocity $v_j$ including BSM 
effects is given by
\begin{align}
	\sigma_j^{\rm BSM}(h_j)v_j=\frac{G_F^2}{2\pi} \abs{V_{ud}}^2\abs{U_{ej}}^2 F_Z(E_e) \, \frac{m_{\rm ^3He}}{m_{\rm ^3H}} E_e \, p_e\, 	T_j(h_j,\epsilon_{lq}),
\label{eq:sigmaNSI}
\end{align}
where $m_{\rm ^3 He}$ and $m_{\rm ^3H}$ are the helium and tritium masses, and $E_e$, $m_e$, 
$p_e$ are the electron e\-ner\-gy, mass and momentum, respectively. The $T_j(h_j,\epsilon_{lq})$ 
function contains the dependence on the neutrino helicity and on the $\epsilon_{lq}$ parameters,
\begin{align}\label{eq:NSIepsilons}
	T_j(h_j,\epsilon_{lq})&={\cal A}(h_j)\left[g_V^2\left(\epsilon_{LL}+\epsilon_{LR}+1\right)^2+3 \, g_A^2\left(\epsilon_{LL}-\epsilon_{LR}+1\right)^2+g_S^2\, \epsilon_{LS}^2+48\, g_T^2\, \epsilon_{LT}^2\right.\notag\\
	&\qquad\qquad\left.+\frac{2m_e}{E_e}\,[g_S\, g_V\, \epsilon_{LS}\left(\epsilon_{LL}+\epsilon_{LR}+1\right)-12\, g_A\, g_T\, \epsilon_{LT}\left(\epsilon_{LL}-\epsilon_{LR}+1\right)]\right]\notag\\
	&\quad +{\cal A}(-h_j)\left[g_V^2\,(\epsilon_{RR}+\epsilon_{RL})^2+3 \, g_A^2\,(\epsilon_{RR}-\epsilon_{RL})^2+g_S^2\, \epsilon_{RS}^2+48\, g_T^2\, \epsilon_{RT}^2\right.\notag\\
	&\qquad\qquad\qquad\left.+\frac{2m_e}{E_e}\,[g_S\, g_V\, \epsilon_{RS}\,(\epsilon_{RR}+\epsilon_{RL})\,-12\, g_A\, g_T \,\epsilon_{RT}\,(\epsilon_{RR}-\epsilon_{RL})]\right]\notag\\
	&\quad+2\,\frac{m_j}{E_j}\left\{g_S\, g_V\,\epsilon_{RS}\,\left(\epsilon_{LL}+\epsilon_{LR}+1\right)+\epsilon_{LS}\,(\epsilon_{RR}+\epsilon_{RL}))\right.\notag\\
	&\qquad\qquad\ \left.-12\, g_A\, g_T(\epsilon_{RT}\,\left(\epsilon_{LL}-\epsilon_{LR}+1\right)+\epsilon_{LT}\,(\epsilon_{RR}-\epsilon_{RL}))\right\}\notag\\
	&\quad+2\,\frac{m_jm_e}{E_jE_e}\left\{g_V^2\,(\epsilon_{LL}+\epsilon_{LR}+1)(\epsilon_{RR}+\epsilon_{RL})+3 \, g_A^2\,(\epsilon_{LL}-\epsilon_{LR}+1)(\epsilon_{RR}-\epsilon_{RL})\right.\notag\\
	&\qquad\qquad\qquad \left.+g_S^2\,\epsilon_{RS}\, \epsilon_{LS}+48\, g_T^2\,\epsilon_{RT}\, \epsilon_{LT}\right\},
\end{align}
\newpage
\noindent with $m_j, E_j$ the mass and energy of the $j$-th neutrino mass
eigenstate and ${\cal A}(h_j) = 1-2h_j v_j$. Let us note that ${\cal
  A}(h_j) \simeq 1 $ for non-relativistic neutrinos, corresponding to
the case on which we will focus in section~\ref{sec:ptolemy}. Furthermore, 
notice that the capture rate is independent of the pseudoscalar couplings $\epsilon_{lP}$. 
The Fermi function $F_Z(E_e)$, which takes cares of the enhancement of the
cross section due to the Coulomb attraction between the proton and
electron, is given by
\begin{align}
	F_Z(E_e)=\frac{2\pi Z \alpha E_e}{p_e\left [1-e^\frac{-2\pi Z \alpha E_e}{p_e}\right]}\, .
\end{align}
Summing over all the neutrino mass eigenstates, one can calculate the total $\rm ^3H$ capture rate
\begin{align}\label{eq:caprate}
	\Gamma_{\CNB}^{\rm BSM} = N_T \sum_{j=1}^3\, \Gamma_{\CNB}^{\rm BSM}(j)= N_T\sum_{j=1}^3\,\esp{\sigma_{j}^{\rm BSM} (+1)\,v_j \,n_{\nu_+^j}+\sigma_j^{\rm BSM}(-1)\,v_j\, n_{\nu_-^j}},
\end{align}
where $N_T$ is the number of nuclei present in the sample and $n_{\nu_\pm^j}$ the number density at the present time of the helical state $\nu_\pm^j$. 
\begin{table}[t]
\begin{center}
	\begin{tabular}{ccc}
		\toprule\toprule
          Form Factor & Value & Reference \\ \midrule
		$g_V(0)$ & $1$ & \protect\cite{Gershtein:1955fb,Feynman:1958ty}\\ 
		$g_A(0)/g_V(0)$ & $1.2646\pm 0.0035$ & \protect\cite{Akulov:2002gh}\\ 
		$g_S(0)$ & $1.02\pm 0.11$  & \protect\cite{Gonzalez-Alonso:2013ura}\\ 
		$g_P(0)$ & $349\pm 9$ & \protect\cite{Gonzalez-Alonso:2013ura}\\
		$g_T(0)$ & $1.020\pm 0.076$ & \protect\cite{Bhattacharya:2015esa}\\ 
		\bottomrule
	\end{tabular}
	\caption{Hadronic form factors considered in this work.}
	\label{tab:HadPar}
\end{center}
\end{table}

\section{Detection of the $\CNB$ by a PTOLEMY-like detector}\label{sec:ptolemy}

A PTOLEMY-like experiment \cite{Betts:2013uya} aims to detect the
$\CNB$ through the neutrino capture by tritium, a reaction that has no
energy threshold.  We can safely assume that $\CNB$ neutrinos are
non-relativistic today\footnote{As we know from oscillation experiments,
 only one neutrino can be massless.} as their root mean momentum is 
$\langle p \rangle\approx 0.6 \ {\rm meV} \ll m_j$~\cite{Weinberg:1962zza}.  
This has two crucial consequences. First, the neutrino flavour eigenstates 
have suffered decoherence into their mass eigenstates, so a detector would, 
in fact, measure the contribution of each neutrino mass eigenstate. Second, 
at the time of the creation of the $\CNB$, i.e.\ when neutrinos decoupled 
from the primordial plasma, they were ultrarelativistic, making chiral and 
helical eigenstates effectively equal. However, as neutrinos evolved into a 
non-relativistic state due to the expansion of the Universe, chirality and
helicity became different. Since neutrinos were free streaming, it was helicity, 
not chirality, that was conserved in the process.\footnote{If neutrinos underwent a
   clustering process, helicity would not be conserved either. We will comment more 
   on this possibility in section~\ref{sec:cosmo}.}
This implies that the neutrino number density is $n_{\nu_+^j}=n_{\nu_-^j}=n_0\approx 56$ cm$^{-3}$ 
in the Majorana case, while $n_{\nu_-^j}=n_0$ and $n_{\nu_+^j}=0$ in the Dirac case. 
If no BSM interactions are present, the function $T_j(h_j,\epsilon_{lq})$ reduces to
\begin{align*}
	T_j(h_j,0)&={\cal A}(h_j)\left[g_V^2+3 \, g_A^2\right],
\end{align*}
from which, using eq.~(\ref{eq:caprate}), we conclude that 
\begin{align}\label{eq:rates_SM}
\Gamma^{\rm M}_{\CNB} = 2 \, \Gamma^{\rm D}_{\CNB}=85.7 \; \rm [kg \, yr]^{-1}\, ,
\end{align}
where $\Gamma^{\rm M}_{\CNB}$ and $\Gamma^{\rm D}_{\CNB}$ are the Majorana and Dirac capture rates. We will consider in section~\ref{sec:cosmo} the modifications to the neutrino abundance due to BSM physics.

The signature of relic neutrinos in a PTOLEMY-like detector is given by the electron created in the capture process.
Nonetheless, tritium can also undergo $\beta$-decay, giving rise to a continuous electron spectrum. As a consequence, one needs to discriminate the electrons produced by the $\CNB$ neutrino capture from the electrons produced by  $\beta$-decays.
Using kinematics, the electrons produced by the $\nu_j$ relic neutrinos capture will have a definite energy \cite{Long:2014zva}
\begin{align}\label{eq:energy_peak}
	E_e^{{\CNB}, j}\simeq m_e+K_{\rm end}^0+2\,m_j,
\end{align}
where $K_{\rm end}^0$ corresponds to the $\beta$-decay endpoint energy. 
This implies that relic neutrinos could produce one or more peaks in the 
electron energy spectrum at energies larger than the endpoint one. 
If so, $\CNB$ and $\beta$-decay events can in principle be discriminated 
from each other. It is clear that the finite energy resolution of the real 
detector plays an essential role in establishing whether the two signals 
can be separated or not. In order to estimate the signal in a more realistic 
way we will follow \cite{Long:2014zva} and convolute the $\CNB$ capture rate 
of eq.~(\ref{eq:caprate}) and the $\beta$-decay background with a Gaussian function
\begin{subequations}
	\begin{align}
	\frac{d\Gamma_{\CNB}^{\rm BSM}}{dE_e}  &= \frac{1}{\sqrt{2\pi\sigma^2}}\sum_{j=1}^{3}\int_{-\infty}^\infty\, dE_e^\prime \,\Gamma_{\CNB}^{\rm BSM}(j) \,\exp\left[-\frac{(E_e^\prime-E_e)^2}{2\sigma^2}\right]\,\delta(E_e-E_e^{{\CNB}, j}), \\
	\frac{d\Gamma_{\beta}}{dE_e}&= \frac{1}{\sqrt{2\pi\sigma^2}}\int_{-\infty}^\infty\, dE_e^\prime \, \frac{d\Gamma_{\beta}}{dE_e^\prime}\,\exp\left[-\frac{(E_e^\prime-E_e)^2}{2\sigma^2}\right],
\end{align}
\end{subequations} 
where $\sigma$ is the expected experimental energy resolution. The complete expression for the $\beta$-decay rate $\displaystyle\frac{d\Gamma_{\beta}}{dE_e^\prime}$ can be found in ref.~\cite{Ludl:2016ane}.

In order to estimate the total number of events produced by the $\CNB$ and $\beta$-decay in the region in which we expect a $\CNB$ signal, we define the full width at half maximum (FWHM) of the Gaussian function as $\Delta = \sqrt{8\ln 2}\, \sigma$. With this definition, we have
\begin{subequations}\label{eq:N_events}
	\begin{align}
		{\cal N}_{\CNB}^{\rm BSM}(\Delta)&=\int_{E_e^{\CNB}-\Delta/2}^{E_e^{\CNB}+\Delta/2}\, dE_e \frac{d\Gamma_{\CNB}^{\rm BSM}}{dE_e},\\
		{\cal N}_{\beta}(\Delta)&=\int_{E_e^{\CNB}-\Delta/2}^{E_e^{\CNB}+\Delta/2}\, dE_e \frac{d\Gamma_{\beta}}{dE_e},
	\end{align}
\end{subequations}
which can be used to define the ratio
\begin{align}
	r_{\CNB}=\frac{{\cal N}_{\CNB}^{\rm BSM}(\Delta)}{\sqrt{{\cal N}_{\beta}(\Delta)}}.
\end{align}
We will consider that the signal can be discriminated from the background 
when $r_{\CNB}\geq 5$. The future PTOLEMY experiment is expected to have 
$\Delta = 0.15$ eV~\cite{Betts:2013uya} in such a way that a single peak
is expected if the sum of the neutrino masses is about $0.1$ eV. For smaller 
masses, a smaller value of $\Delta$ would be needed to discriminate the signal 
from the background. We study more in detail the interplay between $\Delta$, 
neutrino masses and the position of the peaks observed at PTOLEMY-like detectors in Appendix~\ref{sec:ordering}.

\section{On the contributions of BSM physics to $\CNB$ capture rate}\label{sec:bounds}

The BSM lagrangian of eq.~(\ref{eq:lagr}) generates not only new contributions to 
the neutrino capture by tritium, but also modifies other low energy processes. To assess 
the size of the modification to the neutrino $\CNB$ capture rate, we first need to 
take into account the experimental bounds on the $\epsilon_{lq}$ coefficients. Limits 
from Cabbibo Universality~\cite{Hardy:2008gy}, radiative pion decay~\cite{Mateu:2007tr} 
and neutron decays~\cite{Bhattacharya:2011qm} put bounds on the $\epsilon_{Lq}$ left-chiral 
couplings; meanwhile, limits coming from the $\beta$-decay of several nuclei have
been reviewed in ref.~\cite{Severijns:2006dr}. A complete compendium of the limits regarding 
low energy decays is given in refs.~\cite{Cirigliano:2012ab,Cirigliano:2013xha}. For our purposes, 
we will consider the cases considered in ref.~\cite{Severijns:2006dr}, as they include couplings 
with right-handed neutrinos. The constraints are given in terms of the following combinations 
of couplings:
\begin{align}
	\begin{aligned}
		C_V &= g_V(1+\epsilon_{LL}+\epsilon_{LR}+\epsilon_{RL}+\epsilon_{RR}),		& C_V^\prime &= g_V(1+\epsilon_{LL}+\epsilon_{LR}-\epsilon_{RL}-\epsilon_{RR}),\\
		C_A &= -g_A(1+\epsilon_{LL}-\epsilon_{LR}-\epsilon_{RL}+\epsilon_{RR}),		& C_A^\prime &= -g_A(1+\epsilon_{LL}-\epsilon_{LR}+\epsilon_{RL}-\epsilon_{RR}),\\
		C_S &= g_S(\epsilon_{LS}+\epsilon_{RS}),		& C_S^\prime &= g_S(\epsilon_{LS}-\epsilon_{RS}),\\
		C_T &= 4\,g_T(\epsilon_{LT}+\epsilon_{RT}),		&C_T^\prime &= 4\,g_T(\epsilon_{LT}-\epsilon_{RT}).
	\end{aligned}
\end{align}
Accordingly, we need to convert the bounds on the $C_i^{(\prime)}$ into bounds on 
$\epsilon_{lq}$ at 3$\sigma$ C.L. To this end, we have performed a scan over the ranges
\begin{align}
	\begin{aligned}
	-10^{-3}\le & \epsilon_{LL} \le 10^{-3}\, , & 	-10^{-3}\le & \epsilon_{LR} \le 10^{-3} \, ,\\
	-2.8\times 10^{-3}\le & \epsilon_{LS} \le 5\times 10^{-3}\, , & 
	-2\times 10^{-3}\le & \epsilon_{LT} \le 2.1\times 10^{-3} \, ,\\
	\end{aligned}
\end{align}
and
\begin{align}
	\abs{\epsilon_{Rq}}\le 10^{-1},
\end{align}
keeping only the points consistent with each of the allowed regions of
the $C_h^{(\prime)}$ in ref.~\cite{Severijns:2006dr}. Let us notice that, 
to translate the limits into contraints on the $\epsilon_{lq}$ parameters, 
we also scanned over the $g_A(0)/g_V(0)$ value given in table \ref{tab:HadPar} since 
such parameter is affected by the presence of BSM \cite{Gonzalez-Alonso:2013uqa}. 
The ranges in which the scan is performed have been chosen to include the constraints of 
refs.~\cite{Hardy:2008gy,Mateu:2007tr,Bhattacharya:2011qm} in the left-chiral 
coefficients at the 3$\sigma$ level. Although stronger limits can be imposed 
on right-handed couplings using pion decay~\cite{Campbell:2003ir}, we will not 
include them as they are strongly dependent on the flavour structure of the 
model~\citep{Cirigliano:2012ab,Cirigliano:2013xha}. Finally, LHC bounds coming 
from $pp \to e + X + \slashed{E}_T$ have been studied in refs.~\cite{Bhattacharya:2011qm,Cirigliano:2012ab}. 
However, the analysis is performed supposing the interactions of eq.~\eqref{eq:lagr} 
remain pointlike up to the LHC energies, i.e.\ up to a few TeV. To allow for 
the possibility that BSM physics appears just above the electroweak scale, 
in our analysis we will use only the bounds coming from low energy experiments.

\begin{figure}[t]
	\begin{center}
    			\includegraphics[width=0.6\textwidth]{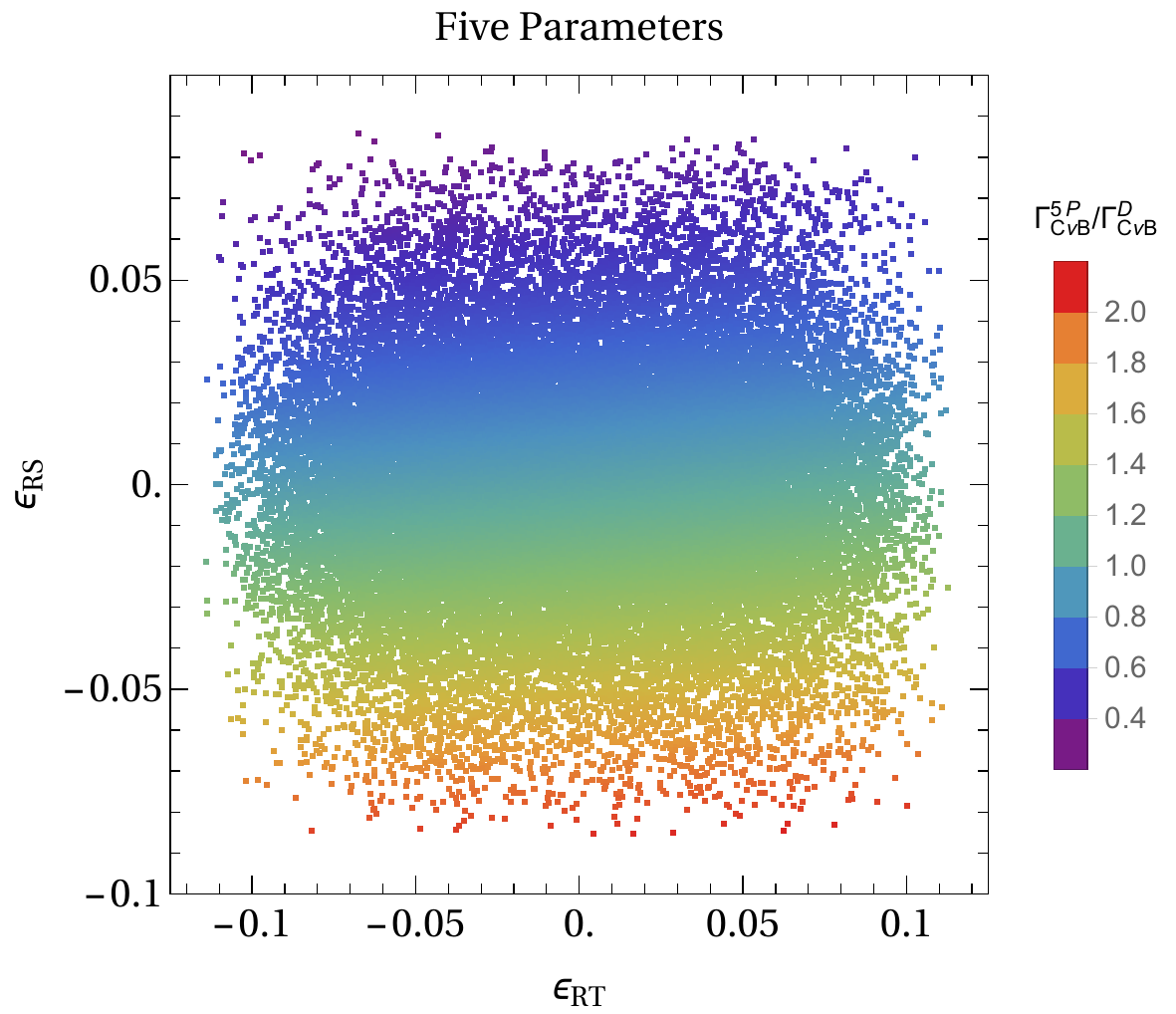}  
	\end{center}
  	\caption{Ratio between the BSM capture rate with respect to the SM Dirac case for the five free couplings scenario  
in the plane  $(\epsilon_{RS}$ versus $\epsilon_{RT})$. The maximum (minimum) value of the ratio is 2.2 (0.3).}
  	\label{fig:5PCase}
\end{figure}

We found that the parameters $\epsilon_{LL}$ and $\epsilon_{LR}$ are unconstrained 
by the experimental data as it has been previously noted in ref.~\cite{Gonzalez-Alonso:2013uqa}.
For reference we summarize here the bounds without the correlations 
--- which have been included in our numerical analysis --- :
\begin{enumerate}

	\item {\it Only left-chiral couplings} allowed in the fit ($\epsilon_{Rq}=0$). 
	The scalar and tensor terms have distinct dependence on the electron energy and mass,
	because of the different Lorentz structure. 
	Computing the total capture rate $\Gamma_{\CNB}^{BSM}$ using the points that pass the low energy experimental constraints, 
	we find 
	\begin{align*}
		0.985 \, \Gamma_{\CNB}^{\rm D} \lesssim \Gamma_{\CNB}^{\rm BSM} \lesssim 1.02 \, \Gamma_{\CNB}^{\rm D}, 
	\end{align*}
	where $\Gamma_{\CNB}^D$ is the capture rate for Dirac neutrinos with only SM interactions.

	\item \emph{Only vector-axial-vector couplings} allowed in the
          fit ($\epsilon_{LS}=\epsilon_{RS}=\epsilon_{LT}=\epsilon_{RT}=0$):
          in this case we get $\vert \epsilon_{RL}\vert \lesssim 8
          \times 10^{-2}$ and $\vert \epsilon_{RR}\vert \lesssim 5
          \times 10^{-2}$ at 3$\sigma$ level. Let us notice that the
          term linear in the right-handed couplings in
          eq.~(\ref{eq:NSIepsilons}) is proportional to
          $m_{j}/E_j$, so it would be negligible for an
          ultrarelativistic neutrino.  This term comes from the
          interference of the SM contribution with the right-handed
          neutrino current. The terms proportional to $(\epsilon
          _{RR}\pm\epsilon_{RL})^2$ come from the square of the
          right-handed currents, and are proportional to ${\cal
            A}(-h_{j})$.  Using the experimentally allowed range for
          $\epsilon_{RR,RL}$, we find 
	  	  \begin{align*}
	    	0.89 \, \Gamma_{\CNB}^{\rm D} \lesssim \Gamma_{\CNB}^{\rm BSM} \lesssim 1.11 \, \Gamma_{\CNB}^{\rm D}.  
	  	  \end{align*}
	  	  
	\begin{figure}[t]
		\begin{center}
    			\includegraphics[width=0.6\textwidth]{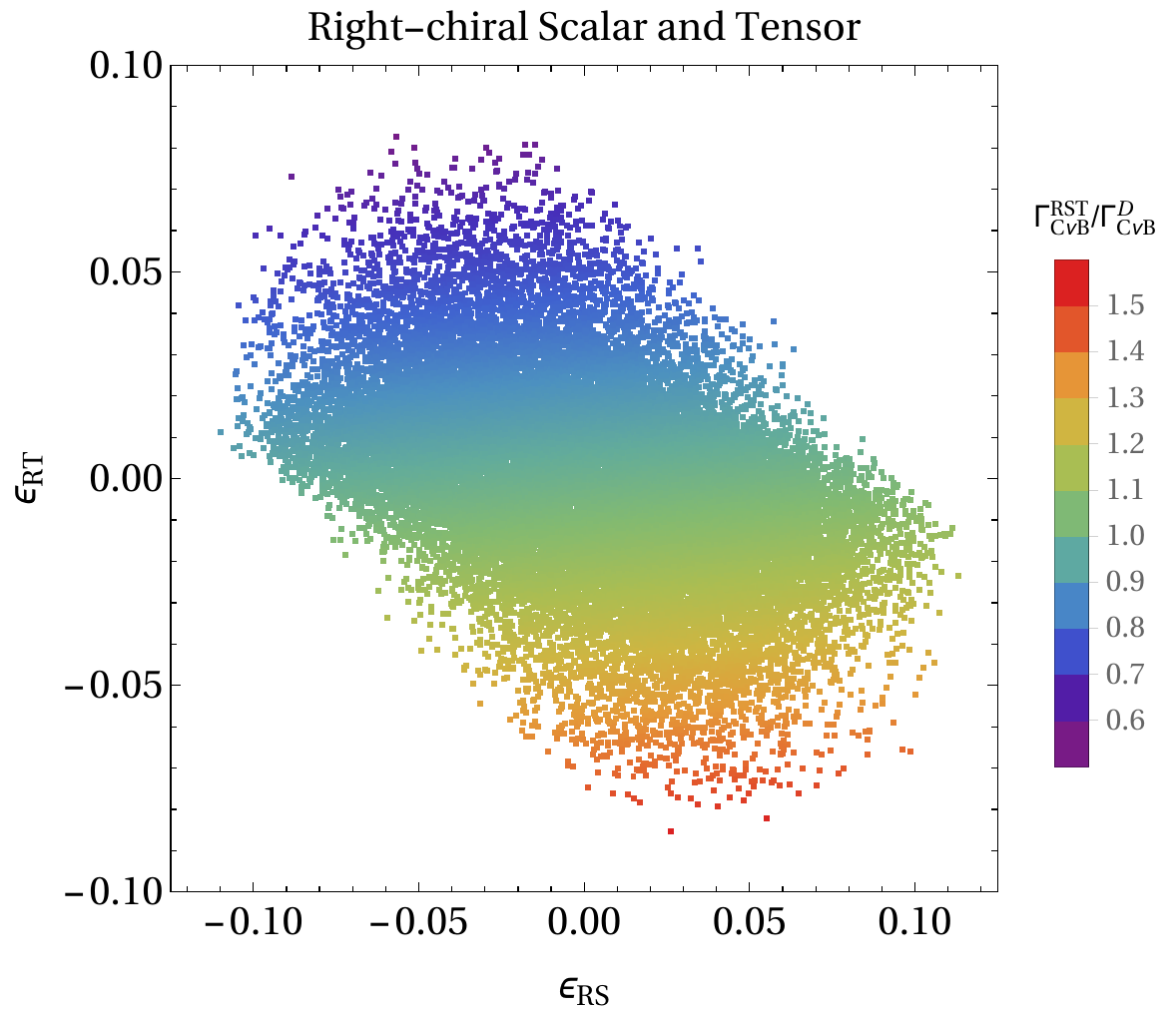}  
		\end{center}
  		\caption{Ratio between the BSM capture rate for the right-chiral scalar and tensor couplings 
  		scenario with respect to the SM Dirac case in the plane $(\epsilon_{RS}$ versus $\epsilon_{RT})$. 
  		We use a color code to indicate the range of values of the ratio.}
  		\label{fig:RSTCase}
	\end{figure}
	  	  
	\item {\it Only right-chiral scalar and tensor couplings}
          allowed in the fit
          ($\epsilon_{LS}=\epsilon_{LT}=\epsilon_{RL}=\epsilon_{RR}=0$):
          in this case we get $\vert \epsilon_{RS}\vert \lesssim 1.1
          \times 10^{-1}$ and $\vert \epsilon_{RT}\vert \lesssim 8
          \times 10^{-2}$ at $3\sigma$. Again the term proportional to the neutrino
          mass comes from the interference between SM and right-handed
          currents. Furthermore, we observe that this interference
          term does not depend on the neutrino helicity. This is due
          to the different Lorentz structures that appear in the BSM
          lagrangian. Considering the allowed parameter space, we
          find
		  \begin{align*}
				  0.61 \, \Gamma_{\CNB}^{\rm D} \lesssim \Gamma_{\CNB}^{\rm BSM} \lesssim 1.52 \, \Gamma_{\CNB}^{\rm D}.  
		  \end{align*}
		  Since in this
          case the parameter space is highly correlated due to the
          correlations coming from the $\beta$-decay bounds, we show
          in figure~\ref{fig:RSTCase} the rate between the BSM capture
          rate and the SM Dirac case in the $(\epsilon_{RS}, \epsilon_{RT})$ plane.

	\item {\it Five free couplings} allowed in the fit: in this
          case we get $\vert \epsilon_{RS}\vert \lesssim 10^{-1}$ 
          and $\vert \epsilon_{RT}\vert \lesssim 8 \times
          10^{-2}$ at $3\sigma$. Here the interference term proportional to the
          neutrino mass depends on the product between $\epsilon_{LS,LT}$
          and $\epsilon_{RS,RT}$.  We show in figure~\ref{fig:5PCase}
          the ratio between the BSM capture rate and the SM Dirac rate
          in the $(\epsilon_{RS}, \epsilon_{RT})$ plane, in which we find the strongest correlation between
          the couplings. We find that the ratio can be at the most 2.2
          times the SM one, which is interesting as in this case Dirac
          neutrinos with BSM interactions can mimic Majorana
          neutrinos in the SM. However, there are regions in parameter space in
          which the rate is considerably lower than the SM one.
\end{enumerate}
Let us conclude stressing that pure Majorana neutrinos fall in the
``only left-chiral couplings" category (case 1 above), with only a small
modification of order $2\%$ allowed in the capture rate. Dirac
neutrinos have instead a much richer phenomenology, with all the above
cases possible (depending on the gauge invariant operators of
table~\ref{tab:d6OPer} generated in the UV theory). On the other
hand, one could also worry about possible modifications of the tritium 
$\beta$-decay spectrum generated by BSM interactions, which could make 
the $\CNB$ detection more involved. Nevertheless, it has been shown in 
ref.~\citep{Ludl:2016ane} that the endpoint of the $\beta$-decay spectrum 
is not significantly modified by BSM physics; thus, in principle, relic 
neutrino detection would be still possible in this case.

\section{On the relic right-handed neutrino abundance}\label{sec:cosmo}

As we have seen in section~\ref{sec:ptolemy}, without BSM contributions 
the neutrino number density today is expected to be
\begin{align}\label{eq:neutr_density}
\begin{aligned}
	n_{\nu_-^j}&=n_0, & n_{\nu_+^j}& =n_0 & &{\rm (Majorana)}, \\
	n_{\nu_-^j}&=n_0, & n_{\nu_+^j}& =0 & &{\rm (Dirac)},
\end{aligned}
\end{align}
with the capture rate in both cases given in eq.~(\ref{eq:rates_SM}). 
There are three ways in which this result can be 
modified: (i) if neutrinos underwent a gravitational clustering process, 
(ii) if BSM interactions are present, and (iii) if an initial abundance
of right-handed neutrinos was present in the early universe.

Neutrino motion in the Dark Matter gravitational potential has the
effect of modifying the direction of the neutrino momentum without
affecting its spin~\cite{Duda:2001hd}. The immediate consequence is
that neutrinos undergo a process of gravitational clustering that
tends to equilibrate the $h_j=+1$ and $h_j=-1$ populations. Since for 
Majorana neutrinos there is already equilibrium, eq.~\eqref{eq:neutr_density} 
is still valid. The situation is different for Dirac neutrinos, for which we get
\begin{align}
	n_{\nu_-^j}&=n_0/2, & n_{\nu_+^j}& =n_0/2 & &{\rm (Dirac, \, clustering)}.
\end{align}
Nevertheless, eq.~\eqref{eq:rates_SM} is still valid since the
additional right-handed neutrino po\-pu\-la\-tion in the Dirac case with
clustering compensates for the loss in the left-handed neutrino
po\-pu\-la\-tion. Very recently, an N-body simulation has been considered 
in ref.~\cite{deSalas:2017wtt} to estimate the relic neutrino density 
enhancement on Earth. The main result is that the clustering effect 
is negligible in the minimal Normal Ordering case while, for minimal 
Inverted Ordering, the capture rate can be increased up to 20\% for 
both Dirac and Majorana neutrinos.

We now turn to the case in which BSM interactions are present. Since
BSM physics modify the electroweak rates, this could potentially
affect the left-handed neutrino abundance. As we have seen in
section~\ref{sec:bounds}, we must have at most $\epsilon_{lq} \lesssim
10^{-1}$ to be compatible with $\beta$-decay and other low energy experimental 
bounds (with many parameters much smaller). As such, the active neutrinos were
maintained in equilibrium with the plasma mainly by SM interactions,
and we do not expect a significant change in the left-handed neutrino number
density $n_{\nu_+^j}$.

Let us finally consider the case in which an initial abundance of right-handed
neutrinos is present. Such abundance can be either thermal or
non-thermal. A thermal population can be achieved by
non-standard interactions or in the presence of a tiny neutrino magnetic 
moment~\cite{Anchordoqui:2012qu,SolagurenBeascoa:2012cz,Zhang:2015wua}. 
Following~\cite{Zhang:2015wua}, when the expansion of Universe becomes faster than the
interaction rate, the right-handed neutrinos decouple as usual. At this freeze out
temperature, $T_R$, the number densities of left- and right-handed neutrinos must
be equal
\begin{align}\label{eq:rhlhab}
	n_{\nu_R^j} (T_R) = n_{\nu_L^j} (T_R).
\end{align}
Using entropy conservation, we can relate the right-handed neutrino
abundance at late times with the left-handed abundance, obtaining~\cite{Zhang:2015wua}
\begin{align}\label{eq:abundance1}
	\frac{n_{\nu_R^j}(T_\nu)}{n_{\nu_R^j}(T_R)}=\frac{g_{*S}(T_\nu)}{g_{*S}(T_R)}\left(\frac{T_\nu}{T_R}\right)^3,
\end{align}
where $g_{*S}(T)$ is the number of relativistic degree of freedom in
entropy at the temperature $T$. Choosing $T_\nu$ in
eq.~\eqref{eq:abundance1} to be the left-handed neutrino decoupling
temperature, and using the definition of the effective number of
thermal neutrino species $N_{\rm eff}$, one
obtains \cite{Anchordoqui:2012qu,SolagurenBeascoa:2012cz,Zhang:2015wua}
\begin{align}\label{eq:RH_abundance_thermal}
	n_{\nu_R^j}(T_\nu) = \left(\frac{1}{3}\Delta N_{\rm eff}\right)^\frac{3}{4} n_{\nu_L^j}(T_\nu),
\end{align}
where $\Delta N_{\rm eff}=N_{\rm eff}^{\rm exp}-3.046$ and $N_{\rm
  eff}=3.046$ is the SM value with 3 left-handed neutrinos. The
experimental determination of $N_{\rm eff}$ by the Planck
collaboration gives~\cite{Ade:2015xua}
\begin{align*}
	N_{\rm eff}^{\rm exp} = 3.14^{+0.44}_{-0.43}\qquad \text{He + Planck TT + low P + BAO}\qquad \text{at $95\%$ C.L.}
\end{align*}
Combining eq.~\eqref{eq:RH_abundance_thermal} with the experimental
result, we get that the maximum density of right-handed neutrinos
is~\cite{Zhang:2015wua}
\begin{align}
	n_{\nu_+^j} &= n_{(\nu_-^j)^c} = n_{0}^R \simeq 16\ {\rm cm}^{-3}.
\end{align}
The relic population of RH neutrinos modifies eq.~\eqref{eq:rates_SM} even
for vanishing non-standard interactions. In the pure SM case, since
the capture rate is proportional to ${\cal A}(h_j)=1$ for both left-
and right-handed neutrinos, we can have an increase in $\Gamma_{\CNB}^{\rm D}$
up to $28 \%$~\cite{Zhang:2015wua}. The difference is even larger if
BSM interactions are turned on, although it depends crucially on the
case considered. For instance, in the vector-axial-vector scenario,
the capture rate is increased by roughly $30 \%$, while in the five
parameter scenario the increase can be up to $70\%$. In this case, we
have that the $\CNB$ rate can be as large as $2.8\,\Gamma_{\CNB}^{\rm
  D}$, reinforcing our results on the possibility of having Dirac
neutrinos with a relic capture rate numerically similar to the
Majorana one.

The last possibility consists in having an initial non-thermal right-handed 
neutrino abundance. Following~\cite{Chen:2015dka}, we will suppose that
right-handed Dirac neutrinos initially form a degenerated Fermi gas, decoupled
from the thermal bath. In this case, the right-handed neutrino density is
related to the photon density $n_\gamma$ by
\begin{align}
	n_{\nu_R^j}(T_\gamma)=\frac{1}{6\zeta(3)}\frac{g_{*S}(T_\gamma)}{g_{*S}(T_R)}\vartheta\, n_\gamma,
\end{align}
where $\vartheta=\varepsilon_F/T_R$, $\varepsilon_F$ the Fermi energy
and $T_R$ the freeze out temperature of the right-handed neutrinos. The
experimental limit on $\vartheta$ obtained using Planck data is
$\vartheta \lesssim 3.26$, from which we get that the maximum right-handed 
neutrino density is~\cite{Chen:2015dka}
\begin{align}
	 n_{\nu_+^j} &= n_{(\nu^j_-)^c} \simeq 36\ {\rm cm}^{-3}.
\end{align}
Since in this case we can have a larger right-handed neutrino
population with respect to the thermal case, we expect larger
modification in the capture rate. In the vector-axial-vector BSM case
we find that the rate is increased between $40$ and $90\%$, getting
closer to the value expected for Majorana neutrinos in the SM. For the
other three scenarios we found larger modifications. In the
right-handed scalar-tensor case, the BSM capture rate has a maximum
value of about $2.5 \, \Gamma_{\CNB}^{\rm D}$, while in the
five-parameter case we obtain $3.5\, \Gamma_{\CNB}^{\rm D}$. We
conclude noticing that, in all the cases in which a right-handed neutrino population
(either thermal or non-thermal) is present, the increase in the number
of neutrinos lead to an increase in the capture rate. 

\section{Conclusions}\label{sec:conc}

The detection of the $\CNB$ would be a milestone for both particle 
physics and cosmology. Experiments using
the neutrino capture in tritium are in development, so that the
detection of the $\CNB$ may become a reality in the near future.  In
this paper we have studied how the capture rate is modified if new
interactions involving neutrinos are present. For definitiveness, we
have focused on the interactions arising from generic BSM physics,
including all the dimension-six operators that can modify the process
$\nu + n \to e + p$. Once the experimental limits coming from low
energy processes are considered, we have seen that for Majorana
neutrinos the modifications to the capture rate are modest (of ${\cal
  O}(2 \%)$), while for Dirac neutrinos we can have much larger
modifications, which can either increase or diminish the capture rate up
to roughly a factor of two. Since in the SM case we expect
the capture rate for Majorana neutrinos to be twice the one for Dirac
neutrinos, we see that the measurement of the capture rate at future
experiments will not be conclusive about the Majorana or Dirac nature
of neutrinos. 

Another situation in which the observed neutrino capture rate can be 
different from the standard one is the existence of a non negligible 
cosmic population of right handed neutrinos. In this case the capture 
rate can either be left unaltered or increase (depending on the physical 
origin of the right handed population). This allows us
to conclude that if a PTOLEMY-like experiment detects a capture
rate \emph{smaller} than the standard capture rate for Dirac neutrinos,
it would unavoidably point to the presence of New Physics in the
neutrino sector (since, as shown in section~\ref{sec:bounds}, the
capture rate can be decreased in this case). If instead the measured
capture rate is between the standard Dirac and Majorana case, or even above 
the standard Majorana case, the situation will not be clear, since the
effect can be caused by Dirac neutrinos with either BSM interactions
or an additional cosmological abundance of right-handed neutrinos. On
the other hand, we have seen how important the right-chiral couplings are
for the relic neutrino capture rate.  Since the rate depends on $
\epsilon_{Rq}$ when $m_j/E_j$ is not negligible, a possible
detection of the $\CNB$ can put stronger limits on the $\epsilon_{Rq}$
couplings that other low energy processes can not.

Finally, we have also briefly discussed in appendix~\ref{sec:ordering} the
problem of distinguishing the electron peaks generated by neutrino capture
and $\beta$-decay. With an expected resolution of $\Delta=0.15$ eV, the PTOLEMY 
experiment will be able to detect only a single peak, corresponding to the capture 
of the three neutrino mass eigenstates. Assuming however two possible resolutions, 
$\Delta=0.01$ eV (very aggressive) and $\Delta=0.001$ eV (ultimate), we established 
a novel criteria to distinguish the electron peaks as a function of the se\-pa\-ra\-tion 
between the experimental Gaussian distributions. The main result is that, given the 
range of neutrino parameters allowed by current oscillation experiments, the ability 
to distinguishing the peaks depends crucially on the neutrino mass ordering, and even 
for the ultimate  value $\Delta = 0.001$ eV the three peaks could be only disentangled 
for normal ordering. This result agrees with previous studies in the 
literature~\cite{Long:2014zva,Blennow:2008fh, Li:2010sn}.

\acknowledgments 

This work was supported by Funda\c{c}\~ao de Amparo \`a Pesquisa do Estado de S\~ao Paulo (FAPESP) and Conselho Nacional de Ci\^encia e Tecnologia (CNPq).

\appendix 

\section{Brief comment on the neutrino mass ordering}\label{sec:ordering}

As we have already stressed, each neutrino mass eigenstate will
produce an electron of energy
given by eq.~(\ref{eq:energy_peak}) in a PTOLEMY-like experiment. A natural question is then
whether each neutrino peak can be distinguished from the $\beta$-decay
background and, if so, when each peak in the distribution can be
distinguished from the peaks generated by the capture of the other
neutrinos~\cite{Blennow:2008fh, Li:2010sn}. The answer depends crucially 
not only on the experimental resolution $\Delta$, but also on the absolute 
value of the neutrino masses as well. In order to answer the above questions, 
we slightly modify eq.~(\ref{eq:N_events}) to consider the number of events due to
the $\nu_j$ capture as
\begin{align*}
{\cal N}_{\CNB}^j(\Delta)&=\int_{E_e^{{\CNB}, j}-\Delta/2}^{E_e^{{\CNB}, j}+\Delta/2}\, dE_e \frac{d\Gamma_{\CNB}^{\rm BSM}(j)}{dE_e}\, ,
\end{align*}
with $E_e^{{\CNB}, j}$ given in eq.~(\ref{eq:energy_peak}). The criteria we use to distinguish the peaks from the background and between each other are the following:

\begin{enumerate}
	\item we say that an electron peak due to neutrino capture can be distinguished from the $\beta$-decay background if  
	\begin{align}
		r_{\CNB}^j \equiv \frac{{\cal N}_{\CNB}^{j}(\Delta)}{\sqrt{{\cal N}_{\beta}(\Delta)}} \geq 5;
	\end{align}
	\item we count the number of distinguishable peaks according to the number of different values taken by the function
	\begin{align}\label{eq:DefXi}
		\Xi^j_{\CNB}=\sum_{i=1}^3\llav{1-\Theta\left(D_B\left(\frac{d\Gamma_{\CNB}^i}{dE_e},\frac{d\Gamma_{\CNB}^j}{dE_e}\right)-4.5\right)}\Gamma_{\CNB}^i ,
	\end{align}
	where $D_B(p,q)$ is the Bhattacharya distance~\cite{MR0010358}, defined for two Gaussians distributions, $p$ and $q$, as
	\begin{align}
		D_B(p,q)=\frac{1}{4}\ln\llav{\frac{1}{4}\corc{\frac{\sigma_p^2}{\sigma_q^2}+\frac{\sigma_q^2}{\sigma_p^2}+2}}+\frac{1}{4}\frac{(\mu_p-\mu_q)^2}{\sigma_p^2+\sigma_q^2}.
	\end{align}
	The value $4.5$, which measures the separation between the
        peaks in the $\Theta$ function of eq.~(\ref{eq:DefXi}), has
        been chosen because it corresponds to a distance of $6\sigma$
        between the mean values of two Gaussians with
        $\sigma_p=\sigma_q$.
\end{enumerate}

\begin{figure}[t]
    \centering 
        \subfloat[$\Delta = 10^{-2}$ eV]{\includegraphics[width=\columnwidth]{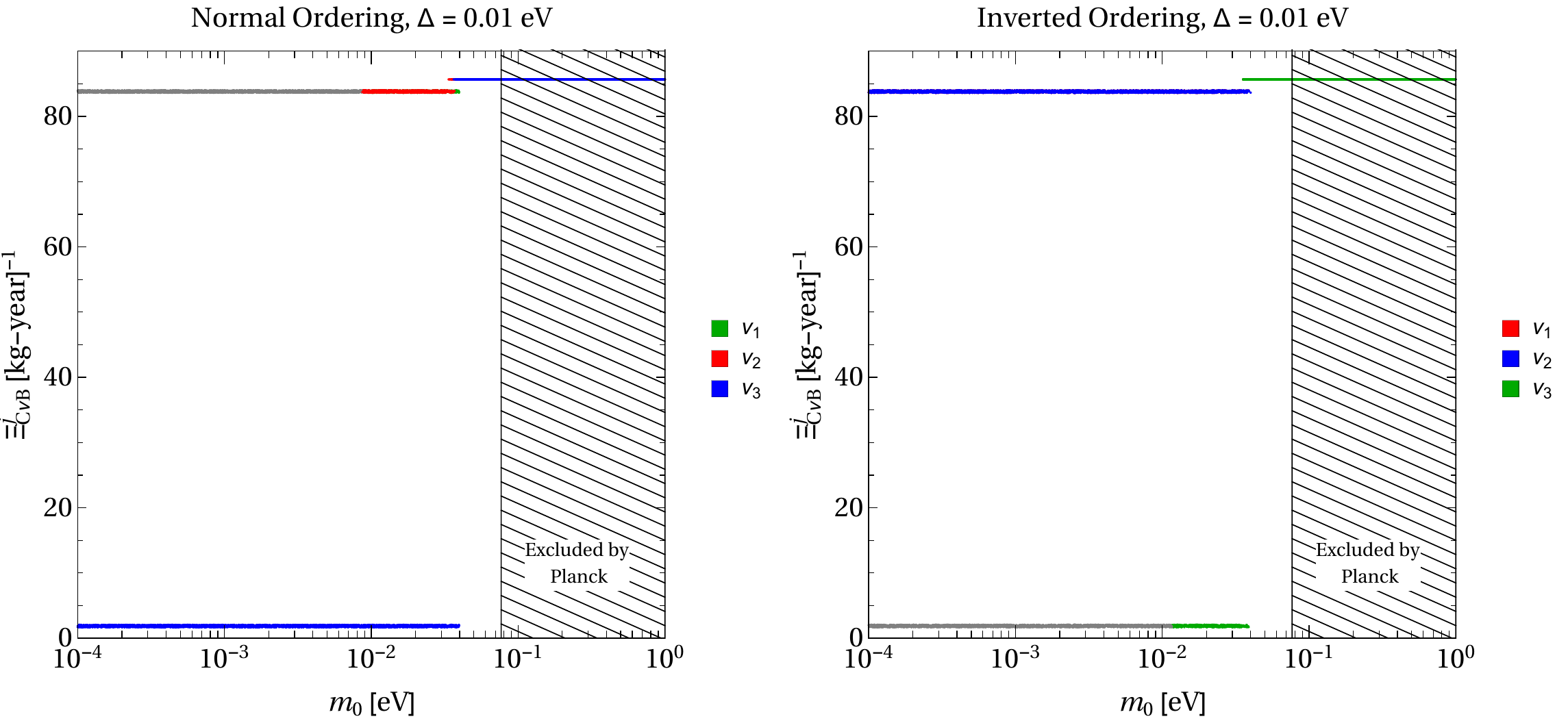}}\\
        \subfloat[$\Delta = 10^{-3}$ eV]{\includegraphics[width=\columnwidth]{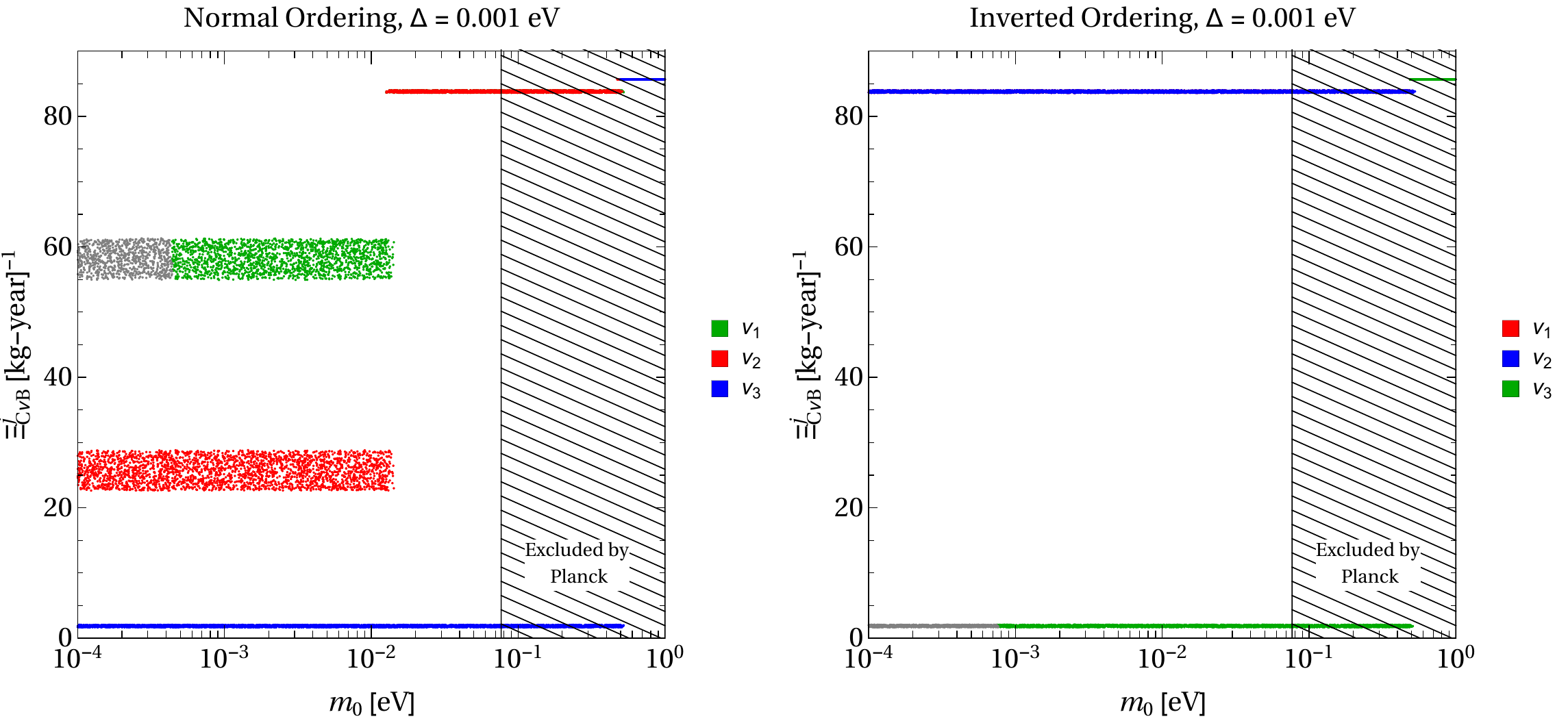}}
        \caption{\label{fig:Xivsm0} Dependence of the $\Xi^j_{\CNB}$
          function of eq.~(\ref{eq:DefXi}) on the value of the lightest neutrino
          mass $m_0$. The experimental resolution is chosen to be $\Delta =
          0.01$ eV (upper panels) and $\Delta = 0.001$ eV (lower
          panels), and we show both normal ordering (left panels) and
          inverted ordering (right panels). The three neutrino mass
          eigenstate contributions are shown in green ($\nu_1$), red ($\nu_2$) and
          blue ($\nu_3$). The gray points correspond to the regions
          that cannot be distinguished from the $\beta$-decay
          background. The shaded region is excluded by the Planck
          limit on the sum of neutrino masses~\cite{Ade:2015xua}.}
\end{figure}

The function $\Xi^j_{\CNB}$ of eq.~(\ref{eq:DefXi}) has been
constructed as follows: when the mass eigenstates are degenerate, the
Bhattacharya distance vanishes and $\Xi^j_{\CNB}$ gives the total
neutrino capture rate. Since $\Xi^j_{\CNB}$ takes a unique value for
the three neutrino states, we have that only one peak will be seen
experimentally. Meanwhile, if any eigenstate is separated enough to
give a distance equal or larger than $6\sigma$, the $\Xi^j_{\CNB}$
will correspond to the value of the capture rate for such mass
eigenstate. Whether a PTOLEMY-like experiment will be able to
distinguish between two or more neutrino capture peaks depends instead
on the mass ordering and on the experimental resolution $\Delta$. With the expected PTOLEMY resolution of
$\Delta=0.15$ eV, the Gaussian peaks for each electron will be too
large to allow a distinction between the different contribution, so
that a unique peak is expected. Nevertheless, we will try to
understand how the electron peaks would look like for better
experimental resolutions, which we take to be $\Delta=0.01$ eV and
$\Delta=0.001$ eV.

\begin{figure}[t]
    \centering
        \subfloat[$\Delta = 10^{-2}$ eV]{\includegraphics[width=0.95\columnwidth]{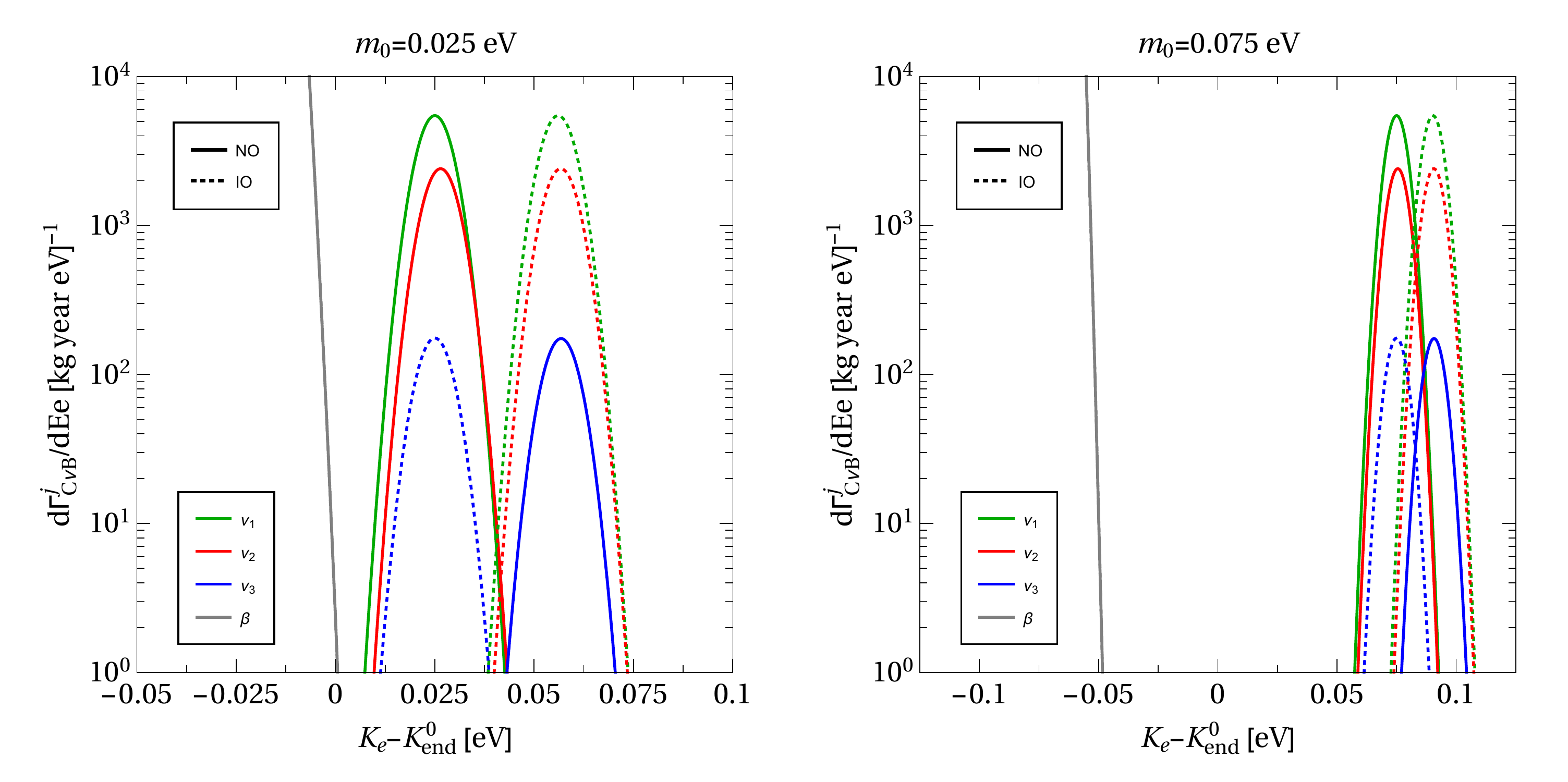}\label{fig:dGammadEe1em2a}}\\
        \subfloat[$\Delta = 10^{-3}$ eV]{\includegraphics[width=0.95\columnwidth]{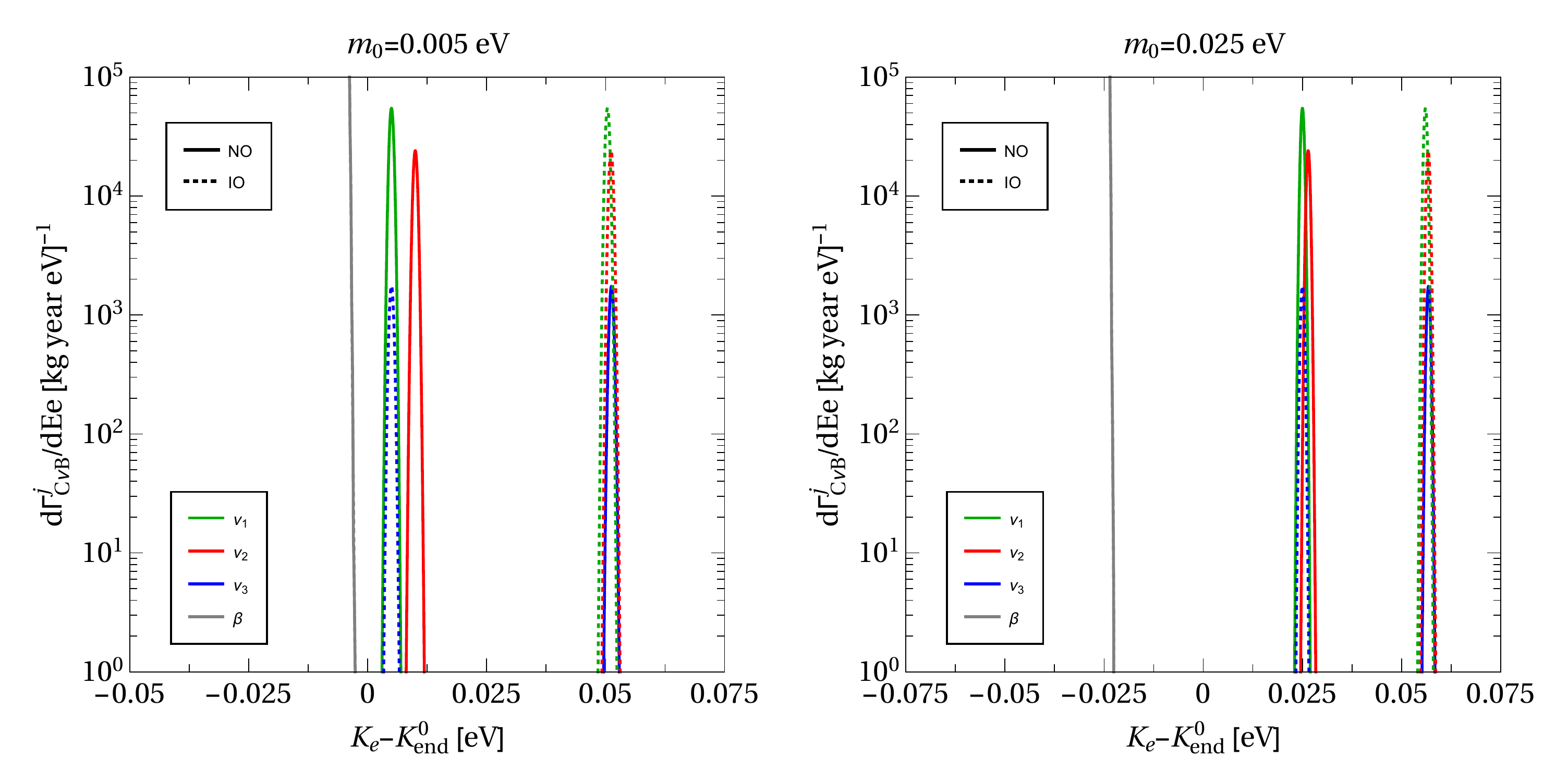}\label{fig:dGammadEe1em2b}}
        \caption{Simulated spectra of the electrons created by the
          relic neutrino capture for $\Delta=0.01$ eV (upper panels) and $\Delta=0.001$ eV (lower panels)
for each mass eigenstate contribution: $\nu_1$ (green), $\nu_2$ (red), $\nu_3$
          (blue). A few values of the lowest neutrino mass $m_0$ are considered to illustrate the behavior for 
the normal ordering (NO) and inverted ordering (IO). The gray line corresponds to the endpoint of the $\beta$-decay
          background.}
  	\label{fig:dGammadEe1em2}
\end{figure}

We show in figure~\ref{fig:Xivsm0} how the $\Xi^j_{\CNB}$ function
depend on the lightest neutrino mass $m_0$, for the mass eigenstates
$\nu_1$ (green), $\nu_2$ (red) and $\nu_3$ (blue). We consider both
types of mass orderings and the two resolution already mentioned,
$\Delta=0.01$ eV and $\Delta=0.001$ eV. We also scan over all the
neutrino parameters at $3\sigma$~\cite{Esteban:2016qun}. The gray
points are those that can not be distinguished from the $\beta$-decay
background. The upper left panel ($\Delta = 0.01$ eV, normal ordering)
should be interpreted as follows: for $m_0 \gtrsim 3 \times 10^{-2}$
eV, the $\Xi^j_{\CNB}$ function takes only one value, so that only one
peak would be measured, which corresponds to the capture of the three
neutrinos. Since the peak is not gray, it can be distinguished from
the $\beta$-decay background. For $8\times 10^{-3}$ eV $\lesssim m_0
\lesssim 3\times 10^{-2}$ eV, two peaks could be measured, one due to
the $\nu_3$ capture (blue) and the other due to $\nu_1$ and $\nu_2$
(red/green). Finally, for $m_0 \lesssim 8\times 10^{-3}$ eV, only the
$\nu_3$ peak can be resolved, while the $\nu_1+\nu_2$ peak cannot be
discriminated from the $\beta$-decay background. The other panels can be
interpreted along the same reasoning. It is interesting to notice that
there is only one situation in which the three peaks can be resolved,
corresponding to the normal ordering for the extreme case $\Delta=0.001$ eV. With the same
resolution but inverted ordering, at most two peaks can be
discriminated, since $\nu_1$ and $\nu_2$ tend to become degenerate as
$m_3 \to 0$.

To better illustrate the interplay between the experimental resolution
$\Delta$ and the importance of the neutrino mass ordering, we show in
figure~\ref{fig:dGammadEe1em2} the expected spectra in a PTOLEMY-like
experiment. In each plot we show normal (continuous line) and inverted
(dashed line) ordering, for the two experimental resolutions we are
discussing (a very agreessive $\Delta = 0.01$ eV, upper panels, and 
an ultimate $\Delta = 0.001$ eV, lower panels) and for some choices for the lightest neutrino mass. The
gray line represents the $\beta$-decay background. This shows another
potential problem in the peak detection; since
\begin{align*}
	\Gamma_{\CNB}^j\propto \abs{U_{ej}}^2,
\end{align*}
and
\begin{align*}
	\abs{U_{ej}}^2 \simeq \left\{0.68, 0.3, 0.02 \right\},
\end{align*}
the peak due to $\nu_3$, although in principle distinguishable from
the other peak(s), is much smaller, and will most probably be unresolved or 
unobservable in a real experiment.

\bibliography{references}{}
\bibliographystyle{JHEP}

\end{document}